\documentclass[a4paper,fleqn]{cas-dc}
    
\usepackage{amsmath,amssymb,amsfonts}
\usepackage{algorithmic}
\usepackage{graphicx}
\usepackage{textcomp}
\usepackage{xcolor}
\usepackage{xspace}
\usepackage{url}
\usepackage{colortbl}
\usepackage{booktabs}
\usepackage{paralist}
\usepackage{algorithm}
\usepackage{balance}
\usepackage{cleveref}
\usepackage{framed}
\usepackage{mdframed}
\usepackage{tabularx}
\usepackage{makecell}
\usepackage{threeparttable}
\usepackage{multirow} 
\usepackage[numbers]{natbib}
\usepackage{diagbox}
\usepackage[many]{tcolorbox}
\usepackage{makecell,multirow,diagbox,rotating}
\usepackage{etoolbox} 
\usepackage{fancybox}

\newcommand{\ie}{\textit{i}.\textit{e}.,\xspace}
\newcommand{\eg}{\textit{e}.\textit{g}.,\xspace}
\newcommand{\DLAP}{DLAP\xspace}

\colorlet{gray1}{gray!70}
\colorlet{gray2}{gray!25}
\colorlet{gray3}{gray!15}
\definecolor{darkgrey}{HTML}{434343}
\definecolor{lightgrey}{HTML}{A9A9A9}
\definecolor{silver}{HTML}{D3D3D3}
\definecolor{midgrey}{HTML}{808080}
\definecolor{white}{HTML}{FFFFFF}
\usepackage{caption}
\newtcolorbox{LYbox}[2][]{text width=0.95\linewidth,fontupper=\normalsize,
fonttitle=\bfseries\sffamily, colbacktitle=darkgrey,enhanced,
attach boxed title to top left={yshift=-2mm,xshift=3mm},
boxed title style={sharp corners},top=4pt,bottom=2pt,left=2pt,right=2pt,
title=#2,colback=white,coltitle=white}

\begin{document}
\makeatletter
\def\verbatim{\small\@verbatim \frenchspacing\@vobeyspaces \@xverbatim}
\makeatother
\renewcommand\arraystretch{1.4}
\let\WriteBookmarks\relax
\def\floatpagepagefraction{1}
\def\textpagefraction{.001}
\shorttitle{A Deep Learning Augmented Large Language Model Prompting Framework}
\shortauthors{Yang et al.}

\title[mode = title]{DLAP: A Deep Learning Augmented Large Language Model Prompting Framework for Software Vulnerability Detection}

\author[a]{Yanjing Yang}

\author[a]{Xin Zhou\corref{cor}}[orcid=0000-0002-3263-1275]
\ead{zhouxin@nju.edu.cn}

\author[a]{Runfeng Mao}

\author[a]{Jinwei Xu}

\author[a]{Lanxin Yang}

\author[a]{Yu Zhang}

\author[b]{Haifeng Shen}

\author[a]{He Zhang}

\address[a]{Sofeware Institute, Nanjing University, China}
\address[b]{Faculty of Science and Engineering, Southern Cross University, Australia}

\begin{abstract}
Software vulnerability detection is generally supported by automated static analysis tools, which have recently been reinforced by deep learning (DL) models. However, despite the superior performance of DL-based approaches over rule-based ones in research, applying DL approaches to software vulnerability detection in practice remains a challenge due to the complex structure of source code, the black-box nature of DL, and the domain knowledge required to understand and validate the black-box results for addressing tasks after detection. Conventional DL models are trained by specific projects and, hence, excel in identifying vulnerabilities in these projects but not in others. These models with poor performance in vulnerability detection would impact the downstream tasks such as location and repair. More importantly, these models do not provide explanations for developers to comprehend detection results. In contrast, Large Language Models (LLMs) have made lots of progress in addressing these issues by leveraging prompting techniques. Unfortunately, their performance in identifying vulnerabilities is unsatisfactory. This paper contributes \textbf{\DLAP}, a \underline{\textbf{D}}eep \underline{\textbf{L}}earning \underline{\textbf{A}}ugmented LLMs \underline{\textbf{P}}rompting framework that combines the best of both DL models and LLMs to achieve exceptional vulnerability detection performance. Experimental evaluation results confirm that \DLAP outperforms state-of-the-art prompting frameworks, including role-based prompts, auxiliary information prompts, chain-of-thought prompts, and in-context learning prompts, as well as fine-turning on multiple metrics.
\end{abstract}


\maketitle

\section{Introduction}
\label{sec: introduction}

Software vulnerability detection is paramount for safeguarding system security and individual privacy. As the cyber environment grows increasingly complex and attack techniques grow quickly, these various threats to software systems have long puzzled software organizations~\cite{lin2020software, steenhoek2023empirical}. In particular, vulnerability is one of the critical threats that may result in information leakage, data tampering, and system breaks~\cite{telang2007empirical}. Vulnerability detection aims to identify vulnerabilities, mitigate their impact, and prevent malicious attacks~\cite{gonzalez2021anomalicious}. Moreover, vulnerability detection helps to enhance software quality, usability, and trustworthiness. Nowadays, vulnerability detection has become a must-have in modern software development.


Many automated static analysis tools (ASATs) have been applied for vulnerability detection~\cite{li2021sysevr, lin2020software}. However, on the one hand, the outputs of ASATs are difficult to validate as they require developers to master more expertise and experience in vulnerability detection~\cite{nachtigall2022large}; on the other hand, the performance of ASATs is poor (\eg high false positive rates) due to they are based on string pattern matching~\cite{christakis2016developers, kang2022detecting}. In recent years, the advancement of deep learning (DL) in natural language processing has inspired researchers to integrate DL models\footnote{In this paper, `DL model' refers to the conventional deep learning models other than large language models such as GPT, Copilot, and Llama.} into ASATs. These modern ASATs generally outperform their conventional counterparts in vulnerability detection~\cite{li2021sysevr, steenhoek2023empirical}. However, DL models that perform well on experimental datasets may suffer from severe performance degradation in real-world projects. This is mainly because of the complexity of source code structure and the concealment of vulnerability characteristics~\cite{chakraborty2022deep}. Using ASATs with DL models to detect vulnerabilities makes an impact on a collection of downstream tasks, including but not limited to vulnerability validation, localization, and repair. Moreover, it is challenging for developers who are responsible for checking vulnerabilities indicated by DL models~\cite{tomas2019empirical}. 

In recent years, Large Language Models (LLMs) such as ChatGPT~\cite{brown2020language} and Copilot~\cite{chen2021evaluating} have shown prominent performance in various tasks~\cite{zhang2023self, li2023enabling, jin2023inferfix}. However, LLMs have not achieved satisfactory results in vulnerability detection~\cite{purba2023software}. Dai et al.~\cite{dai2023can} indicated that one of the main reasons is the inappropriate use of LLMs. LLMs are pre-trained by a vast amount of data, but not all of them have positive effects on the downstream tasks such as vulnerability detection~\cite{hsieh2019classification, liu2023not}. There are two techniques to address this problem: \emph{fine-tuning} and \emph{prompt engineering}. Fine-tuning is a commonly used technique but requires significant computational resources and time. LLMs with prompts allow users to interact with LLMs iteratively to produce bespoke results~\cite{wei2022chain, dai2023can}. However, as detection performance is highly susceptible to prompts, a generic prompting framework would not be able to achieve satisfactory performance~\cite{arakelyan2023exploring}. Prompt engineering makes LLMs adapt to a specific downstream task and generate customized outputs~\cite{wei2022chain, dai2023can}. Moreover, prompt engineering can jointly work with fine-tuning, making it a cost-effective and promising technique for vulnerability detection~\cite{shi2023large}.  

Previous work has utilized LLMs for vulnerability detection using various prompting frameworks~\cite{zhang2023prompt, dai2023can, purba2023software, lu2024grace}. However, existing prompts input limited information to LLMs, making them provide little help in improving the performance of LLMs in real-world projects. To address this problem, we propose a bespoke prompting framework \DLAP\footnote{Data and materials: 
\url{https://github.com/Yang-Yanjing/DLAP.git}}. 
Although DL models can not achieve unsatisfactory performances in multiple projects, they have superior performance in a single project. The core idea of \DLAP is using pre-trained DL models for the target project to stimulate adaptive implicit fine-tuning of LLMs. We select the most suitable DL model as plugins among three categories to augment \DLAP. This process is implemented through two state-of-the-art prompts: In-context Learning (ICL) prompt and Chain-of-Thought (COT) prompt. On the one hand, the ICL prompt utilizes locality-sensitive hash (LSH) to sample candidate code fragments that are similar to the input code. Then, a pre-trained DL model is employed to obtain the prediction probabilities of candidate fragments. The combination of the candidate code fragments and their corresponding probabilities forms the ICL prompt for the input code. On the other hand, the COT prompt synthesizes the results from static scanning tools and pre-trained DL models as queries. Following these queries, \DLAP locates the corresponding COT templates within the detection step template library we constructed based on Common Weakness Enumeration (CWE\footnote{\url{https://cwe.mitre.org}}). Then \DLAP uses these COT templates to generate the customized completed detection COT prompts for input codes. This stimulates LLMs to conduct implicit fine-tuning to achieve better performance in vulnerability detection and provide supplementary information to facilitate the inspection and comprehension of detection results.

We conduct experiments using four large-scale projects with more than 40,000 examples to evaluate \DLAP. We first conduct experiments to select the most suitable DL model to form \DLAP. We assess performance by integrating various DL models into \DLAP and comparing their results to determine the optimal deep learning model. The results show that combining Linevul with an LLM outperformed other DL models by 15\% across all evaluation metrics. Then we select the Linevul to generate prompts within \DLAP and compare \DLAP against the state-of-the-art prompting frameworks including role-based prompts, auxiliary information prompts, chain-of-thoughts prompts, and in-context learning prompts. The results show that \DLAP  surpasses baselines across all metrics for each project, achieving a 10\% higher F1 score and a 20\% higher Matthews Correlation Coefficient (MCC). This indicates that \DLAP is more effective for vulnerability detection. Finally, we compare our approach with the most prevalent fine-tuning techniques to explore the effectiveness of \DLAP versus fine-tuning. The results reveal that \DLAP can achieve 90\% of the extensive fine-tuning process at a lower cost and even outperform fine-tuning on some metrics. Moreover, the \DLAP-driven LLM model generates more explanatory text than fine-tuning, which is important to aid developers in using ASATs for vulnerability detection tasks.

The main contributions of this paper are as follows.

\begin{itemize}
    \item We propose \DLAP, a bespoke LLM prompting framework for vulnerability detection. \DLAP combines the advantages of DL models and LLMs while overcoming their respective shortcomings. Additionally, \DLAP has the potential to be adapted for other ASAT tasks.
    \item We conduct rigorous experiments to demonstrate the effectiveness of selecting appropriate DL models for \DLAP and showcase the exceptional vulnerability detection performance over state-of-the-art prompting frameworks.
    \item We empirically demonstrate the advantages of prompting over fine-tuning for vulnerability detection in terms of detection accuracy, cost-effectiveness, and explanations.  
    
\end{itemize}

The rest of the paper is organized as follows. \Cref{sec:related} reviews the background and related work. \Cref{sec: framework} delineates the design of \DLAP. \Cref{sec:experiments} presents the experimental design and parameter setting of \DLAP, followed by results and analysis in \Cref{sec:analyis}. \Cref{sec:discussion} discusses \DLAP's generation capability and DL model selection. Finally, we present threats to validity in \Cref{sec:ttv} and conclude this paper in \Cref{sec:conclusion}.

\section{Background and Related Work}
\label{sec:related}
This section describes the related work on vulnerability detection and the background of prompt engineering for LLMs.

\subsection{Vulnerability Detection}
While there is a plethora of work on this topic, we focus on vulnerability detection enhanced by DL and LLMs.

\subsubsection{Deep Learning for Vulnerability Detection}
Vulnerability detection has received a lot of concerns in recent years. Lin et al.~\cite{lin2020deep} proposed a framework that incorporates one project and 12 DL models for slice-level vulnerability detection. Zhou et al.~\cite{zhou2019devign} proposed Devign which uses graph representation for input performed better than approaches using tokens of code directly. Li et al. developed a series of DL-based approaches including VulDeePecker, $\mu$VulDeePecker, and SySeVR~\cite{li2021sysevr,li2018vuldeepecker,zou2019mu} to complete the construction of a DL framework applied to vulnerability detection. Despite achieving advanced results in experimental setups, there still exists generalization issues in practical applications. With the advent of networks based on the transformer architecture and language models, researchers have started applying these advanced NLP techniques to vulnerability detection. Fu and Tantithamthavorn~\cite{fu2022linevul} applied RoBERTa as a pre-training model, fine-tuned on subsequent vulnerability detection tasks, achieving the best experimental performance in both function-level and line-level vulnerability prediction tasks. 

Chakraborty et al.~\cite{chakraborty2022deep} found that the performance of several DL-based approaches dropped by an average of 73\% on datasets built from multiple real-world projects, highlighting the need for further research into cross-project vulnerability detection. Steenhoek et al.~\cite{steenhoek2023empirical} conducted an empirical study to demonstrate the variability between runs of a model and the low agreement among DL models' outputs and studied interpretation models to guide future studies. 

\subsubsection{Large Language Model for Vulnerability Detection}
The outstanding performance in dialogue, code generation, and machine translation of LLMs has sparked the interest of researchers and practitioners in applying LLMs to software security. Katsadouros et al.~\cite{katsadouros2023can} highlighted the potential of LLMs to predict software vulnerabilities, emphasizing their advantage over traditional static methods. Thapa et al.~\cite{thapa2022transformer} discovered that transformer-based LLMs outperformed conventional DL-based models. Zhang et al.~\cite{zhang2023prompt} enhanced the effectiveness of ChatGPT in software vulnerability detection through innovative prompt designs and leveraging the model's ability to memorize multi-round dialogue. 

However, according to Cheshkov et al.~\cite{cheshkov2023evaluation}, ChatGPT and GPT-3 in Java code vulnerability detection do not outperform the current tools. In the meantime, Liu et al.~\cite{liu2023not} emphasized that ChatGPT cannot replace professional security engineers in vulnerability analysis, indicating that closed-source LLMs are not the end of the story. These findings suggest that the performance of LLMs in the realm of vulnerability detection leaves much to be desired. The potential false positives and illusions generated by LLMs in specific applications \cite{zhang2023siren} are attributable to the extensive unconstrained training data and the multitude of training parameters. Consequently, it is essential to fine-tune an LLM before deploying it for specific tasks. Lu et al.~\cite{lu2024grace} proposed a method called GRACE that processes code structure using CodeT5, combines semantic with syntactic features to conduct similarity searches, and utilizes in-context learning prompts to drive the LLM beyond all baseline DL models on complex real-world datasets. 

As indicated above, the performance of LLMs remains unsatisfactory, accompanied by a high false positive rate. In this study, GRACE~\cite{lu2024grace} along with other evaluated prompts~\cite{zhang2023prompt} will serve as the baselines, facilitating the comparison and evaluation. To achieve better performance in vulnerability detection, we select Sysver~\cite{li2021sysevr}, Devign~\cite{zhou2019devign}, and Linevul~\cite{fu2022linevul} as the augmented components for our framework.

\subsection{Prompt Engineering for Large Language Models}
The increase in the number of parameters in LLMs leads to a rise in the cost of fine-tuning LLMs. Low-cost approaches such as Low-Rank Adaptation of Large Language Models (LoRA)~\cite{hu2021lora} and P-tuning~\cite{liuetal2022p} have significantly reduced the cost of fine-tuning. However, the cost for some applications is still considerable. For example, fine-tuning an LLM with 33B parameters requires two high-precision 40G GPUs~\cite{hu2021lora}. The parameters of the LLMs that have been reported to achieve excellent performance all exceed one hundred billion, which may cost a lot of computing resources. Researches~\cite{brown2020language, chowdhery2022palm, bai2022constitutional} reveal that LLMs are transformer-based models. Different inputs can cause changes in the attention layers within their architecture, and as such, the construction of high-quality prompts can assist the LLMs in providing satisfactory answers for the specific target tasks. In contrast, inappropriate prompts will impinge on its attention, which may mislead LLMs to produce hallucinations~\cite{zhang2023siren}. The COT~\cite{wei2022chain} and ICL~\cite{dai2023can} prompting are currently the most effective approaches to prompt engineering. COT prompting is an approach of decomposing a target problem into steps to prompt LLMs to provide the answers. ICL prompts LLMs to deliver correct answers by referring to similar questions. In this paper, \DLAP integrates both approaches to provide appropriate prompts to drive LLMs for vulnerability detection.

\begin{figure*}[htbp]
    \begin{center}
    \includegraphics[width=0.9\textwidth]{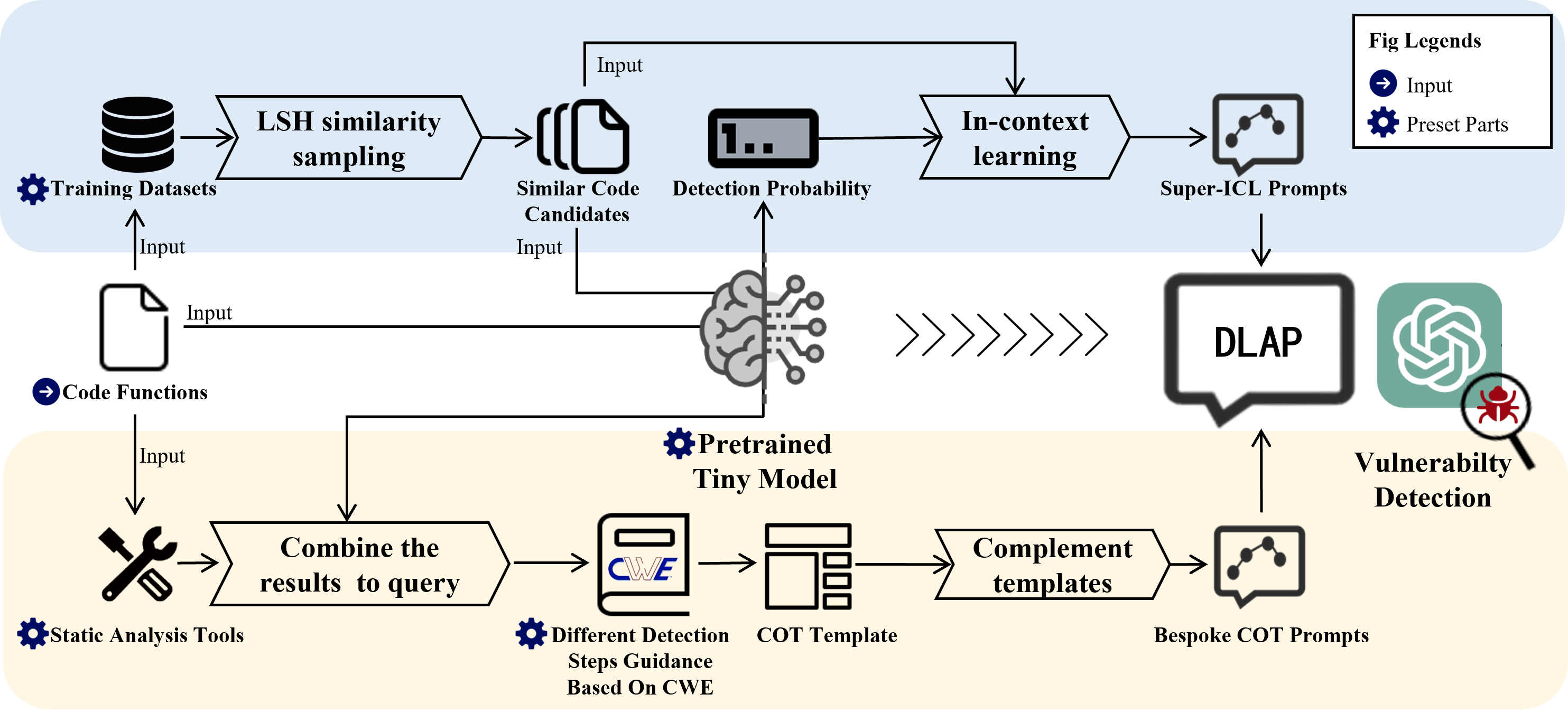}
    \vspace{-2.0ex}
    \end{center}

    \caption{An overview of the proposed \DLAP}
    \vspace{-4.0ex}
\label{fig: Overview}
\end{figure*}

\section{The Proposed Framework: \DLAP}\label{sec: framework}
This section describes the design of \DLAP, which consists of two prompt techniques.

\subsection{Motivation}
\label{sec: motivation}
Vulnerability detection can be formulated as a binary classification problem. Given a vulnerability dataset $\mathcal{D}$ that is represented with $\mathcal{D} = \{(x_i, y_i)\}_{i=1}^N, y=\text{0 or 1}$, where $x_i$ is the source code of a function, and $y_i$ is the ground-truth (0--NO, 1--YES). Detection models are expected to establish their mappings automatically. One of the main processes of detection models is input (source code) representation. Specifically, the source code can be represented as semantic tokens, abstract parsing trees (ASTs), data/control flow graphs (DFGs/CFGs), or other formats. Conventional DL models use only a single format as input, which may miss useful information. LLMs have the capability of combining multiple representations, making it a more promising detection technique for vulnerability detection. When levering LLMs for vulnerability detection, it is fundamental to make LLMs understand domain and task knowledge as LLMs are trained by general corpora. It can be expected that using LLMs for vulnerability detection directly would achieve unsatisfactory results. Therefore, LLMs use \emph{fine-tuning} or \emph{prompt engineering} to address this task. 

Fine-tuning is an intuitive technique to make the parameters $\mathcal{W}$ of LLMs adapt to downstream tasks (\ie vulnerability detection) ion $\mathcal{D}_{o}^E$ to achieve better results, which can be described as \Cref{eq:motivation-FT}.

\begin{equation}
\footnotesize
\hspace{1.5cm}
    \begin{aligned}
        \mathcal{W} = \text{argmin}\mathcal{L}\left(Y-\mathcal{M}_\mathcal{W}(\mathcal{P}(X))\right)
    \end{aligned}
\label{eq:motivation-FT}
\end{equation}
where $Y$ is the ground truth (label) for function level code and $\mathcal{L}$ is the loss function. By fine-tuning the weight parameters $\mathcal{W}$ of LLMs, the predictive probabilities are expected to be closer to the ground truth. However, fine-tuning is cost-intensive~\cite{hu2021lora,white2023prompt}. For example, LoRA which is one of the most efficient LLMs requires approximately 80G graphics card memory and a lot of time for fine-tuning an LLM with only 13B parameters.

Prompt engineering is a new technique to augment LLMs. LLMs can incorporate various inputs and generate their answers; therefore, they can be prompted ~\cite{brown2020language, chowdhery2022palm, bai2022constitutional}. Technically, we use $\mathcal{D}_{s}^T$ and $\mathcal{D}_{o}^E$ to describe pretraining sets and testing sets, respectively. When using prompt engineering ($\mathcal{P}(\dot)$ ) for vulnerability detection, LLMs accept a set of $\mathcal{P}(X)$ as inputs and output the estimated probabilities of them, where $X$ means a collection of examples from $\mathcal{D}_{o}^E$. One cost-effective prompting template $\mathcal{P}$ for vulnerability detection can be described as \Cref{eq:motivation-PE}. 

\begin{equation}
\footnotesize
\hspace{1.5cm}
    \begin{aligned}
        \mathcal{P} = \text{argmin}\mathcal{L}\left(Y-\mathcal{M}_\mathcal{W}(\mathcal{P}(X))\right)
    \end{aligned}
\label{eq:motivation-PE}
\end{equation}

According to Liu et al.~\cite{liuetal2022p}, \Cref{eq:motivation-PE} can achieve the same effects as \Cref{eq:motivation-FT}. That is, both prompt engineering and fine-tuning can make predictive probabilities close to ground truths. In the following subsections, we elaborate on \DLAP which leverages prompt engineering for vulnerability detection.

\subsection{Framework Overview}
\DLAP leverages attention mechanisms within LLMs, incorporating selectively trained DL models as enhancements. This approach, known as In-Context Learning (ICL), acts to subtly refine LLMs, making them more adept at specific projects. Moreover, \DLAP's use of chain-of-thoughts (COT) enables LLMs to discard incorrect generative paths effectively. Consequently, \DLAP enhances LLMs' capabilities in detection tasks, ensuring robust performance without incurring significant costs. ICL can stimulate the attention layer of an LLM to adapt to the downstream detection task, which is defined by~\cite{dai2023can} as implicit fine-tuning. As with general fine-tuning, implicit fine-tuning can also drive LLMs to adapt to downstream tasks and achieve better performance. Well-designed prompts stimulate LLMs to perform better in downstream detection tasks. The idea behind the proposed \DLAP framework is that it uses DL models to augment LLMs by constructing appropriate prompts to stimulate implicit fine-tuning of the LLMs for them to adapt to vulnerability detection tasks. In this way, it can reduce performance degradation caused by hallucination and data distribution differences. 

As shown in~\Cref{fig: Overview}, \DLAP is composed of two main parts, including (1) Part I: construction of in-context learning prompts augmented by DL models and (2) Part II: generation of the bespoke COT prompts to augment LLMs. In Part I, we employ DL models to generate detection probabilities for input codes and select candidate codes based on similarity. The combination of candidate codes and their corresponding similarities forms the ICL prompt for detection. In Part II, we combine the results of DL models and static tools to query pre-defined templates in a preset COT template library as key-value pairs. Based on the characteristics of each input sample, we complete the chain of thought, generating COT prompts for detection. These two parts will be introduced in \Cref{sec: framework-ICL} and \Cref{sec: framework-COT} respectively. In \Cref{sec: Prompts Synergy}, we show an example of synergizing the two prompts to generate the prompts of the final DLAP.

\subsection{In-Context Learning Prompts Construction}
\label{sec: framework-ICL}
According to the earlier point, LLMs encapsulate vast knowledge through their expansive weight structures.
Firstly, we select the pre-trained DL model through training sets. For new projects, we can also build DL models from newly collected samples based on existing research~\cite{fu2022linevul,li2021sysevr,zhou2019devign}. Then to create the appropriate in-context, the most similar code candidates are found in the training set using Locality-sensitive hashing (LSH), an efficient similarity search algorithm in the Retrieval Augmented Generation (RAG) technique. Although the LSH similarity calculation algorithm can only concern the similarity of code segments, considering that as a prompting framework, we cannot spend too much time generating prompts formation, it is necessary to sample multiple codes as a similar code candidate set efficiently.

Following Dai et al.~\cite{dai2023can}, we reversely use the dual form of the attention of transformer derived by them. Therefore, the adaptive implicit fine-tuning on attention layer $\widetilde{\mathcal{A}}$ of LLMs stimulated by \DLAP for specific projects can be written as \Cref{eq softatten}. Please refer to \Cref{proveICL} in the appendix for details.
\begin{equation}
\hspace{1.2cm}
\footnotesize
 \begin{aligned}
 \widetilde{\mathcal{A}}(\mathbf{q}) =  ( W_{\text{init}} + \Delta W_{ICL}(x)) \mathbf{q}
    \end{aligned}
\label{eq softatten}
\end{equation}

We train the DL model $\mathcal{M}$ with training data information $\texttt{info}$ which contains the relationship between the project data and the label. Then the DL model generates a detection probability $\operatorname{Probs_\mathcal{P}}$ for a detection object $x$ as shown in \Cref{DL formulas}.
\begin{equation}
\hspace{1.2cm}
\footnotesize
\begin{aligned}
\operatorname{{Probs_{ICL}}}(Obj_\text{info}) = \mathcal{M}(Obj_\text{info})(x)
\end{aligned}
\label{DL formulas}
\end{equation}

The probabilities output by the DL model represent characteristics of input codes.
\DLAP uses the probabilities to construct ICL prompts to augment LLMs.
Then, we obtain the relaxed attention representation $\widetilde{\mathcal{A}}$ of the LLM in \Cref{eqIFT}.

\begin{equation}
\footnotesize
\hspace{0.8cm}
 \begin{aligned}
 \widetilde{\mathcal{A}}(\mathbf{q})= \left( W_{\text{init}} + \Delta W(\operatorname{func}(\operatorname{Probs_\mathcal{P}}(Obj_\text{info}))) \right) \mathbf{q}
    \end{aligned}
\label{eqIFT}
\end{equation}

\Cref{eqIFT} indicates that the relaxed attention of the LLM is related to probabilities output by the selection of a DL model and the probabilities are related to the training project data. This results in an implicit fine-tuning for the LLM to adapt to specific project information~\cite{dai2023can}. We further explain this adaptive process with more details in the result analysis section through a comparative analysis experiment.

\begin{figure}[htbp]
\begin{center}
    \includegraphics[width=1.0\linewidth]{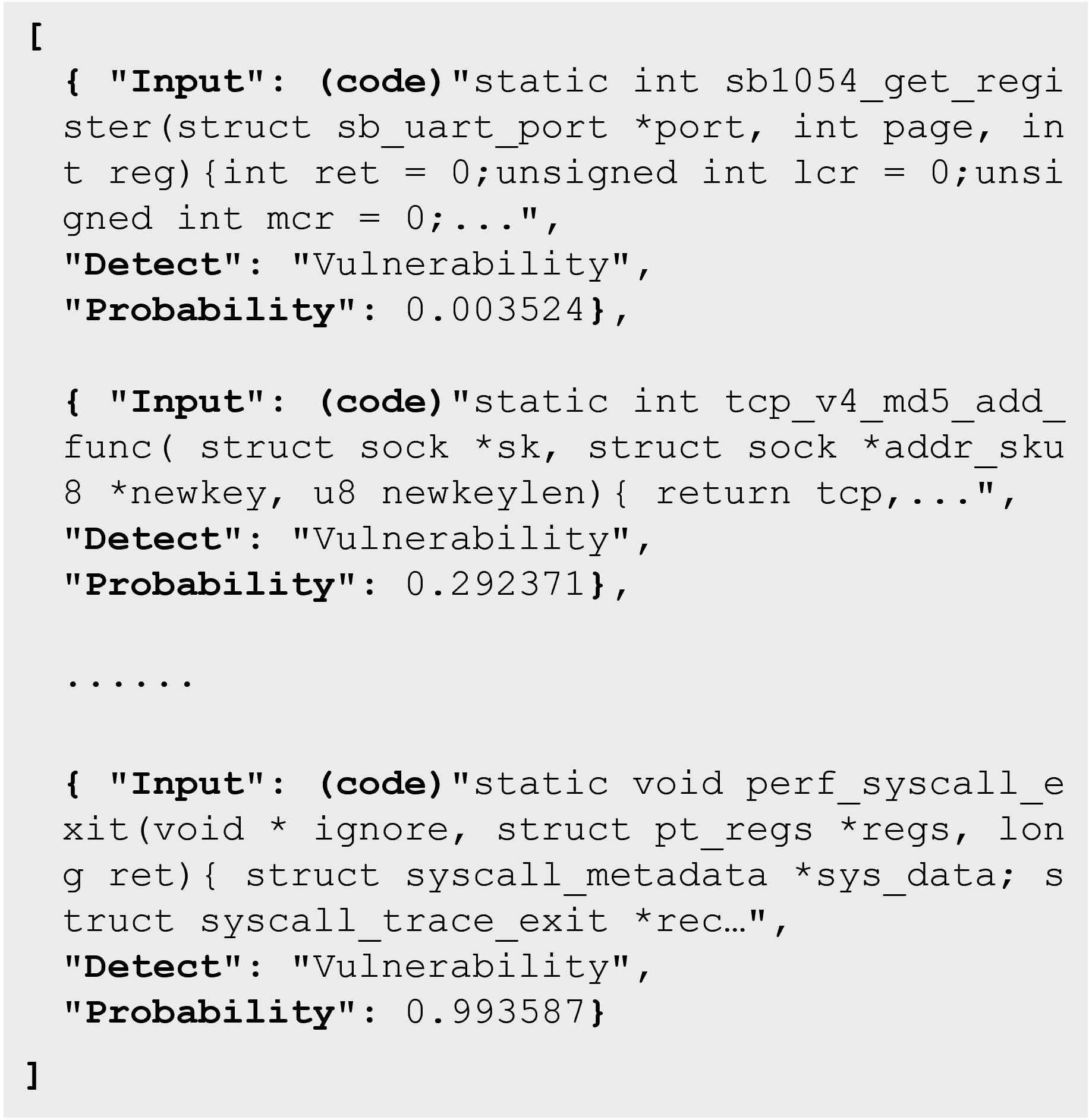}
    \vspace{-2.0ex}
\end{center}
\vspace{-4.0ex}
\caption{An example of ICL prompts of \DLAP}
\vspace{-4.0ex}
\label{fig: ICLexample}
\end{figure}

Compared to conventional fine-tuning methods, \DLAP does not require excessive resource consumption to update the parameters of LLMs. The ICL prompts update the output of the attention layer in the LLMs. As the example shown in \Cref{fig: ICLexample}, the ICL prompts of \DLAP stimulate implicit fine-tuning of the LLMs toward adapting to the characteristics of the projects to be detected.  

By adding the results generated by the DL model to the prompts of LLM, \DLAP can include more in-context information. The DL model trained to form weights of networks contains the complex relationship between the predicted probability and the input code text. 
Therefore the DL model's output contains the characteristics of training sets. 
ICL built in this way contains more prompt information than normal ICL, which stimulates LLMs to perform better in downstream tasks.

As shown in the upper part of \Cref{fig: Overview}, after conducting the detection probability of the candidate codes through the DL model, we set the code candidate set and the corresponding probability as the question and answer combination. These combinations (an example shown on \Cref{fig: ICLexample}) are constructed to be the in-context learning (ICL) prompt together with part II in \Cref{sec: framework-COT} to form the final \DLAP augmented prompts.

\subsection{Chain-Of-Thought Prompts Generation}
\label{sec: framework-COT}

The second part of \DLAP is to generate specific prompts for each tested sample. It is divided into the following stages. Firstly, because the characteristics and detection steps of vulnerabilities vary, we need to pre-set different detection templates into a COT library. According to the existing peer-reviewed vulnerability taxonomies (\ie \cite{wei2021comprehensive, li2017novel}) and reliable grey literature~(\ie \cite{tsipenyuk2005seven}), we construct a hierarchical detection COT library that has six major categories as follows.

\begin{itemize}
    \item \textbf{SFE} (Security Features Errors): Errors induced by imperfect security features
    
    \item \textbf{LOG} (Logistics Errors): Errors induced by program execution

    \item \textbf{MEM} (Memory Errors): Errors related to memory resources

    \item \textbf{NUM} (Numeric Errors): Errors induced by numerical computations
    
    \item \textbf{IDN} (Improper Data Neutralization): Errors induced by non-standardization (verification, restriction) of exchanged data
    
    \item \textbf{UNT} (Unknown Taxonomy Errors): Unknown errors
\end{itemize}
Subsequently, according to the parent-child relationship described in the CWE research concept\footnote{\url{https://cwe.mitre.org/data/definitions/1000.html}}, some categories above are refined (a total of 45 subcategories). 
In addition, by referring to the relevant research~\cite{liu2023not, zhang2023prompt,  ozturk2023new} on step-by-step solutions to vulnerability detection through LLMs driven by the COT, we establish a general paradigm for the generation of the COT as follows.
\begin{itemize}
\item \textbf{Semantics}: 
Comprehending the function of the code. 
\item \textbf{Logic}:
Analyzing the structure of the code. 
\item \textbf{Internal risks}:
Identifying components that may introduce vulnerabilities. 
\item \textbf{External risks}:
Inspecting for unsafe functions that could potentially lead to vulnerabilities. 
\item \textbf{Generating the COT}:
Integrating the information acquired above and generating a COT to inquire about whether there are potential vulnerabilities step by step. 
\end{itemize}

\begin{figure}[htbp]
\begin{center}
    \includegraphics[width=1.0\linewidth]{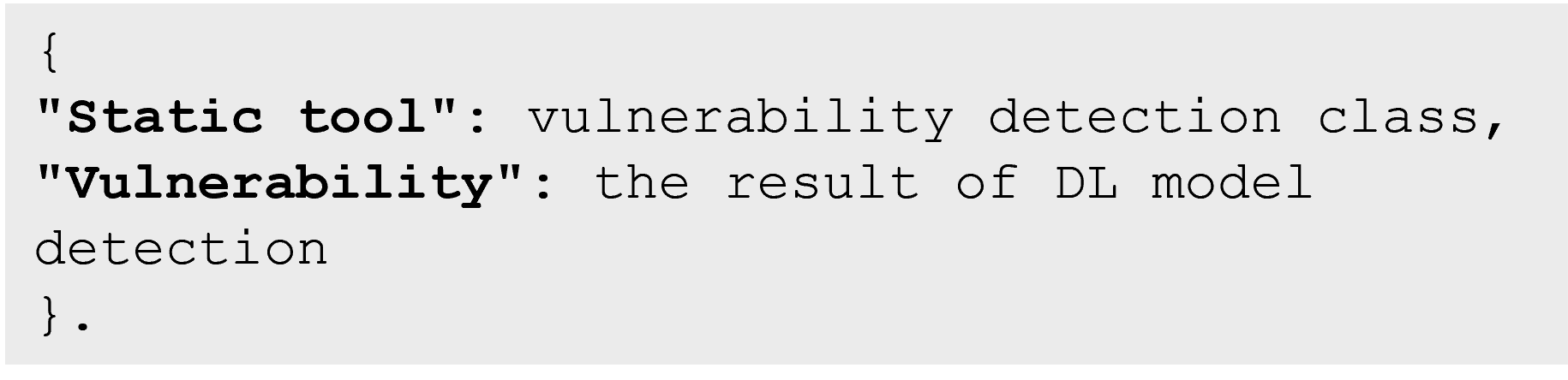}
    \vspace{-2.0ex}
\end{center}
\vspace{-4.0ex}
\caption{A query key example of \DLAP}
\vspace{-2.0ex}
\label{fig: key}
\end{figure}

The specific COT refines the generation paradigms for corresponding COT guidance for different categories. Each subcategory is associated with a specific detection COT template guidance. \DLAP selects two open-source functional-level vulnerability detection static tools, \ie Flawfinder\footnote{\url{https://dwheeler.com/flawfinder}} and Cppcheck\footnote{\url{http://cppcheck.net}}, to generate static scanning results. It parses the result text of the static tools and maps them to the corresponding categories in the taxonomy tree. Then the results are scored and recorded for each tool. The highest-scoring K categories are taken out and added to the query \textit{key}. 
\DLAP selects the same DL model in \Cref{sec: framework-ICL} because of their better performance according to the studies~\cite{zhou2019devign, fu2022linevul,li2021sysevr}. 
\DLAP combines the detection results of the DL model and the scanning results of the static tool to become the \textit{key} of a query. Using this \textit{key}, \DLAP obtains customized COT generation guidance templates from the COT library for the test codes. The \textit{key} is formed as a dictionary that contains the static tool output class and the result of the DL model judgment, such as the following \Cref{fig: key}

\begin{figure}[htbp]
\begin{center}
    \includegraphics[width=1.0\linewidth]{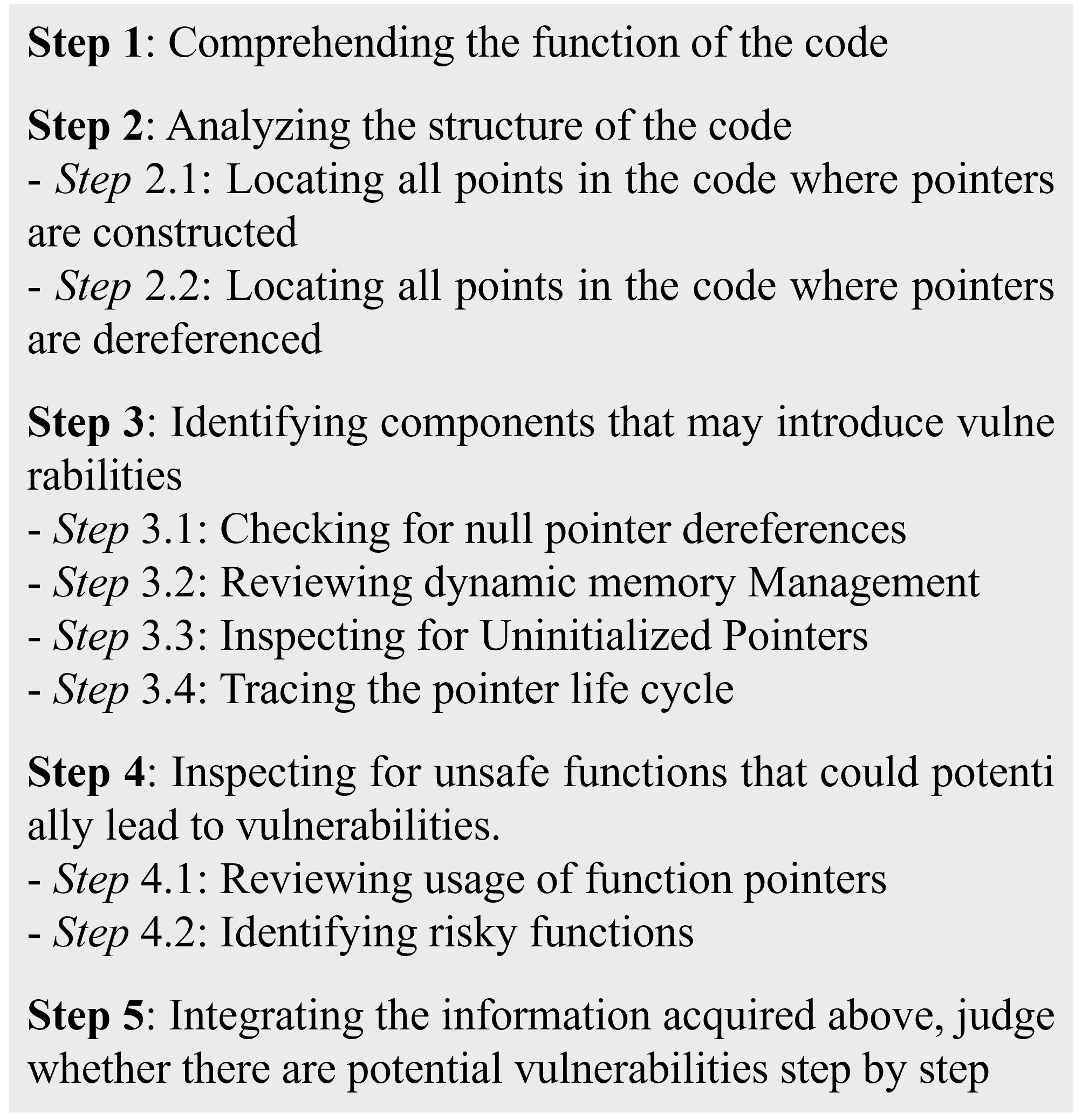}
    \vspace{-2.0ex}
\end{center}
\vspace{-4.0ex}
    \caption{An example of the refined COT prompts of \DLAP}
    \vspace{-2.0ex}
\label{fig: COTexample}
\end{figure}

For instance, if the \textit{key} is the null pointer dependency that falls under the IDN category, then \DLAP will get the refined COT guidance from the taxonomy tree as shown in \Cref{fig: COTexample}. The library of COT guidance templates is publicly available on GitHub\footnote{\url{https://github.com/Yang-Yanjing/DLAP.git}\texttt(COTTree)}.

Through the \textit{key} generation process described earlier, the COT guidance and the results of the DL model are combined to generate final COT prompts for the target-specific detection samples. 
\begin{figure*}[htbp]
\begin{center}
    \includegraphics[width=1.0\linewidth]{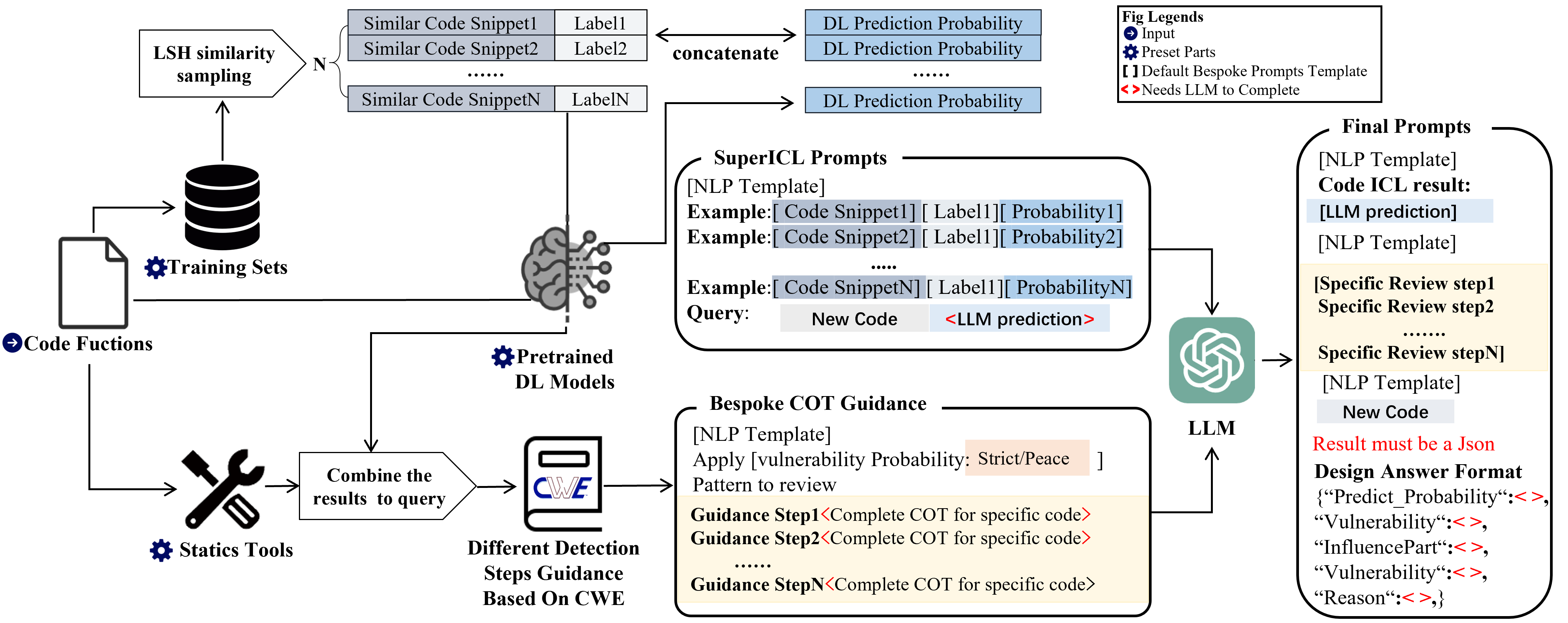}
    \vspace{-2.0ex}
\end{center}
\vspace{-4.0ex}
    \caption{An example of the refined COT prompts of \DLAP}
    \vspace{-4.0ex}
\label{fig: example}
\end{figure*}

\subsection{Prompts Synergy}
\label{sec: Prompts Synergy}

\Cref{fig: example} shows an example of our algorithm where the final prompts are made in such a format.
The process of \DLAP generating prompts is described in \Cref{algorithm: DLAP}, which takes the selected DL models, the training/history data of target detection project $\mathcal{X}=\{(x_i, y_i)\}_{i=1}^N$, the selected static tools $\mathcal{S}$, and the preset COT library $\mathcal{S}$ as the input. 

\begin{algorithm}[!ht]
\caption{Bespoke prompts for detection samples}
\label{algorithm: DLAP}
\fontsize{9.5}{12}\selectfont
\begin{algorithmic}[1] 
\REQUIRE $\mathcal{M}$; $\mathcal{X}=\{(x_i, y_i)\}_{i=1}^N$; $\mathcal{S}$; $\mathcal{K}$.
\STATE $\mathcal{X}_{o}^T\leftarrow   \operatorname{Sample}(\mathcal{X})$  \# the training set for constructing the DL-based model\\

       $Y\leftarrow \operatorname{Label}(\mathcal{X}_{o}^T)$\\
       $\mathcal{X}_{o}^E=\mathcal{X} - \mathcal{X}_{o}^T $
\STATE $\mathcal{M} = \operatorname{argmin}\mathcal{L}(\mathcal{X}_{o}^T, Y)$\\
\FOR{$x_i,i=1$ to $N$ in $\mathcal{X}_{o}^E$}
    \STATE $[Candidates] = LSH(x_i)$
    \STATE $[Probabilities] = \mathcal{M}([Candidates])$
    \STATE ICL prompt$(x_i) = \{[Probabilities],[Candidates]\}$ 
    \STATE $Result_\mathcal{S}(x_i)\leftarrow \operatorname{Ranking}(\mathcal{S}(\mathcal{X}_{o}^E))$
    \STATE $\mathcal{P}(x_i)= Prediction_\mathcal{M}(x_i)+Result_\mathcal{S}(x_i)$
    \STATE Utilizing $\mathcal{P}(x_i)$ as the key to retrieve COT prompts generation guidance. $G(x_i) = \mathcal{K}(\mathcal{P}(x_i))$
    \STATE Using GPT to complete $G(x_i)$: $COT(x_i)=GPT(G(x_i))$
    \STATE \DLAP prompts$(x_i)$ = ICL prompt$(x_i)$ + COT prompt$(x_i)$
\ENDFOR
\ENSURE Specific COT prompts of all detection samples.
\end{algorithmic}
\end{algorithm}

First, the target detection project $X$ is sampled to construct training sets $\mathcal{X}_{o}^T$. $\mathcal{X}_{o}^T$ (if open-source information is directly collected) is labeled as $Y$. The remaining part of detecting project $X$ is test sets $\mathcal{X}_{o}^E$. $\mathcal{X}_{o}^T$ and label $Y$ are utilized to minimize the loss function $\mathcal{L}$ for training $\mathcal{M}$. Next, for each input code in the test set $\mathcal{X}_{o}^E$, the LSH algorithm is used to find the most similar code $[Candidates]$ in this set. Then, the DL model $\mathcal{M}$ is utilized to get detection probabilities $[Probabilities]$. Using $[Candidates]$ and $[Probabilities]$, \DLAP constructs question-answering combinations to form the DL-augmented ICL prompts.

\DLAP carries out a bespoke process described in \Cref{fig: example} to get the specific COT prompt. Statics tools are utilized to get analysis results $\mathcal{S}(\mathcal{X}_{o}^E) = \{\mathcal{S}_1:\text{scores}, \mathcal{S}_2:\text{scores}, \mathcal{S}_3:\text{scores}, ... \}$ and $Result_\mathcal{S}(x_i)$ is obtained by ranking $\mathcal{S}(\mathcal{X}_{o}^E)$. Then \DLAP generates the DL model prediction result. After that, the results $Predictions_\mathcal{M}$ are combined to form a query, which is utilized as the key to retrieve COT prompts generation guidance $G(x_i)$ from the COT library. GPT is used to complete $G(x_i)$ for generating $cot(x_i)$ for each detection sample. Finally, the final prompts of \DLAP are composed of the COT prompts and ICL prompts. Each specific prompt is utilized to drive LLMs for detection. The final step is to use the generated COT prompts for LLMs to produce understandable vulnerability detection results by following a particular answer format.

\section{Experimental Design}\label{sec:experiments}
This section details research questions, datasets, DL models, baseline LLM prompts, and evaluation metrics.

\subsection{Research Questions}
The experimental evaluation of \DLAP is structured with three research questions (RQs). 

\smallskip
\begin{itemize}
    \item [\textbf{RQ1:}] \textbf{Which category of DL models is the most effective to DLAP?}
\end{itemize} 

\noindent\emph{\textbf{Motivation\&Setup:}}
An important driver of \DLAP is the DL model, which adds the information stored by the DL model training to the prompting process of LLMs through the ICL approach. As little is known about which category of DL models is suitable for augmenting LLMs in the context of vulnerability detection, we propose RQ1 to compare three representative DL models: Sysevr, Devign, and Linevul (cf. \Cref{ex: DL models} for rationales). 

\smallskip
\begin{itemize}
    \item [\textbf{RQ2:}] \textbf{How effective is \DLAP compared to existing prompting frameworks?} 
\end{itemize} 

\noindent\emph{\textbf{Motivation\&Setup:}}
Previous research has shown that the performance of LLMs is susceptible to prompts and inappropriate prompts lead to unsatisfactory performance. In this paper, we design \DLAP as a prompt augmented framework for vulnerability detection. In RQ2, we compare \DLAP against four existing prompting frameworks: \textbf{P}$_{\text{Rol}}$, \textbf{P}$_{\text{Aux}}$, \textbf{P}$_{\text{Cot}}$, GRACE (cf. \Cref{ex: baselines} for rationales) to evaluate its effectiveness.


\smallskip
\begin{itemize}
    \item [\textbf{RQ3:}] \textbf{How effective is \DLAP compared to LoRA fine-tuning?} 
\end{itemize}

\noindent\emph{\textbf{Motivation\&Setup:}}
Previous research has shown that fine-tuning is helpful for augmenting LLMs. In this paper, we use prompt engineering rather than fine-tuning to develop \DLAP because of its lower cost. In RQ3, we compare \DLAP against a fine-tuning LLM (Llama-13B)~\cite{touvron2023llama} to see if it has the same performance as fine-tuning. Specifically, we select LoRA~\cite{hu2021lora}, a state-of-the-art LLM fine-tuning technique for comparison.



\subsection{Datasets}
\label{ex: datasets}
According to Croft et al.~\cite{croft2023data}, the common vulnerability detection datasets have labeling bias. To develop an appropriate experimental dataset, we customize three criteria for selecting projects: (1) it has been researched by related work~\cite{chakraborty2022deep, zhou2019devign, fan2020ac} (to ensure external validity); (2) it has accumulated more than 3,000 functions (to exclude no-active projects); and (3) it is traceable (to exclude projects whose vulnerability information is incorrect or even unknown). As a result, our experimental dataset consists of four open-source projects, including Chrome, Linux, Android, and Qemu. These projects we selected are of good open-source quality and have high-quality vulnerability fix records for traceability.

The basic information of the selected projects is shown in \Cref{tab: dataset}, from which we can observe the number of trues (vulnerabilities) and falses in each project is imbalanced. To mitigate the impact of data imbalance on training the DL augment model, we first performed random undersampling on the non-vulnerable samples of the four projects. Then we divided the dataset into training and testing sets with the 8:2 proportion. The training set was used to build DL models, while the testing set was used to evaluate the performance of DLAP.
\begin{table}[htbp]
\scriptsize
\centering
\renewcommand\arraystretch{1.25}
\caption{\textbf{Basic information of dataset}}
\label{tab: dataset}
\begin{threeparttable}
\begin{tabular}{c | ccc}
\hline
\textbf{Project}& \textbf{\#Functions}& \textbf{\#Vulnerbilities}& \textbf{Used by}\\
\hline
\hline
\textbf{Chrome}& 77,173 & 3,939 & Chakraborty
et al.~\cite{chakraborty2022deep} \\
\textbf{Linux}& 46,855 & 1,961 & Fan et al.~\cite{fan2020ac} \\
\textbf{Android}& 8,691 & 1,277 & Fan et al.~\cite{fan2020ac} \\
\textbf{Qemu}& 3,096 & 125 & Zhou et al.~\cite{zhou2019devign} \\
\hline
\end{tabular}
\vspace{-0.0ex}
\end{threeparttable}
\end{table}

\subsection{DLAP Refinement}
\label{ex: DL models}
To address RQ1, we select three DL models for vulnerability detection to refine \DLAP. Each of the three represents one type of DL model. Their rationales and hyperparameter settings are as follows.

\begin{itemize}
    \item \textbf{Sysevr}~\cite{li2021sysevr} represents the category that uses code features including syntactic, semantic, and vector representation. It filters the code into slice input by static analysis of semantics and syntax.
    \item \textbf{Devign}~\cite{zhou2019devign} represents the category that introduces more structured graph structures and graph neural networks into the vulnerability detection model.
    \item \textbf{Linevul}~\cite{fu2022linevul} represents the category that utilizes pre-trained deep learning model. This novel system detection model is based on the Transformer architecture.
\end{itemize}

\begin{table}[htbp]
\scriptsize
\centering
\renewcommand\arraystretch{1.25}
\caption{\textbf{The settings of DL model}}
\label{tab: hyperparas}
\begin{threeparttable}
\begin{tabular}{lll}
\hline
\textbf{DL \newline Model}& \textbf{Hyperparameter}& \textbf{Selection}\\
\hline
\multirow{13}{*}{\textbf{Sysevr}}
& Java version & Java 8 \\
& Static tools & Joern 0.3.1 \\
& Graph database & Neo4j \\
& Data preprocessing & Slice \\
& \textbf{Embedding algorithm} & \textbf{Word2vec} \\
&~~~-sampling algorithm & CBOW  \\
&~~~-sampling window & 5 \\
&~~~-min\_Count & 5 \\
& \textbf{Network architecture} & \textbf{BiLSTM}  \\
&~~~-epoch & 100 \\
&~~~-batch\_size & 32 \\
&~~~-optimizer & sgd \\
&~~~-loss function & binary cross-entropy \\
\hline
\multirow{15}{*}{\textbf{Devign}}
& Java version & Java 8 \\
& Static tools & Joern 2.0.157 \\
& Data preprocessing & Graph \\
& \textbf{Embedding algorithm} & \textbf{Word2vec} \\
&~~~-verctor\_size & 100\\
&~~~-epoch & 10 \\
&~~~-min\_count & 1\\
& \textbf{Network architecture} &\textbf{CNN}  \\
&~~~-epoch & 200 \\
&~~~-batch\_size & 128\\
&~~~-input\_channels & 115 \\
&~~~-hidden\_channels & 200 \\
&~~~-num\_of\_layers & 6 \\
&~~~-optimizer & adam \\
&~~~-loss function & binary cross-entropy \\
\hline
\multirow{7}{*}{\textbf{Linevul}}
& Data preprocessing & Slice \\
& Embedding algorithm & BPE+Transformer \\
& \textbf{Pretrained model} & \textbf{codeBERT} \\
&~~~-batch\_size & 256\\
&~~~-num\_attention\_head & 12 \\
&~~~-optimizer & Adam \\
&~~~-loss function & binary cross-entropy \\
\hline
\end{tabular}
\vspace{-0.0ex}
\end{threeparttable}
\end{table}

We selected these three DL models to evaluate a range of similar models for the categories they each represent. As part of our model selection process, we referenced the parameters reported in the respective research papers of these DL models that achieved the best performance. These parameters were selected in \Cref{tab: hyperparas} as the pre-set hyperparameters in our framework. By doing so, we aim to replicate the optimal performance achieved by these models and ensure consistency in our evaluation and comparison.

\subsection{Baselines}
\label{ex: baselines}
We compare \DLAP against four prompting frameworks~\cite{zhang2023prompt, purba2023software, lu2024grace, white2023prompt} that leverage LLMs to detect vulnerabilities.

\begin{enumerate}[0]
\item[$\bullet$ \textbf{P}$_{\text{Rol}}$](Role-based prompts): According to White et al.~\cite{white2023prompt}, providing GPT with a clear role would greatly alleviate its illusion problem. Our first baseline is proposed by Zhang et al.~\cite{zhang2023prompt}, making GPT a vulnerability detection system.
\begin{LYbox}{\textbf{\textit{Role-based prompts}}}
I want you to act as a \textbf{Vulnerability Detection System}. 
My first request is “Is the following program buggy?” Please answer Yes or No. [CODE]
\end{LYbox}

\item[$\bullet$ \textbf{P}$_{\text{Aux}}$](Auxiliary information prompts): Based on the view of Zhang et al.\cite{zhang2023prompt}, providing the LLMs more semantic information about the code for vulnerability detection improves its performance. Therefore, in baseline 2, we provide data flow as auxiliary information to prompts.
\begin{LYbox}{\textbf{\textit{Auxiliary information prompts}}}
I want you to act as a vulnerability detection system. I will provide you with the original program and the data flow information, and you will act upon them. Is the following program buggy? [CODE], [DF description].
\end{LYbox}

\item[$\bullet$ \textbf{P}$_{\text{Cot}}$](Chain-of-thought prompts): According to Wei et al.\cite{wei2022chain}, due to the potential capabilities of LLMs for multi-turn dialogue, constructing a COT better assists LLMs in reasoning~\cite{wei2022chain}. Therefore, in baseline 3, we constructed a two-step thinking chain to drive the LLM in the process of vulnerability detection.
\texttt{Step1}: To make LLMs correctly determine whether a code is vulnerable. This step drives the LLMs to understand the purpose of the code exactly. Therefore, we designed first-step prompts for detecting the intent of the code.
\texttt{Step2}: Based on the first step, we continue to prompt LLMs to detect vulnerabilities for inputs.
\begin{LYbox}{\textbf{\textit{Chain-of-thoughts prompts}}}
\texttt{Step1:}Please describe the intent of the given code. [CODE].\newline
\texttt{Step2:}I want you to act as a vulnerability detection system. Is the above program buggy? Please answer Yes or No
\end{LYbox}

\item[$\bullet$ \textbf{GRACE}:] GRACE is a vulnerability detection prompting framework that enhances the capabilities of LLM for software vulnerability detection. It achieves this by incorporating graph structural information from the code. GRACE employs codeT5 and ICL techniques to use graph information.
\end{enumerate}

\subsection{Evaluation Metrics}
As vulnerability detection is formulated as a binary classification problem in this paper, we use precision ($\mathcal{P}_\text{vul}$), recall ($\mathcal{R}_\text{vul}$), and F1-score ($F_1$) to measure the performance of each framework. Considering vulnerability is a minor class but is of great severity, we also use FPR as a metric. FPR pays attention to false positives since making mistakes on them would cause more serious outcomes than making mistakes on false negatives. In this paper, the minor class (positive) is vulnerability, and it occupies a very small portion. The definition of FPR is shown in \Cref{evalfpr}. Moreover, Matthews correlation coefficient (MCC) is also used as an evaluation metric. MCC, a.k.a., phi coefficient, is a metric to measure the performance of binary classifiers on imbalanced datasets. MCC is a more comprehensive metric than FPR. The definition of MCC is shown in \Cref{evalmcc}.

\begin{equation}
\footnotesize
\hspace{1.9cm}
\text{FPR} = \frac{\text{FP}}{\text{FP} + \text{TN}}
\label{evalfpr}
\end{equation}

\begin{equation}
\footnotesize
\text{MCC} = \frac{{\text{TP} \times \text{TN} - \text{FP} \times \text{FN}}}{{\sqrt{{(\text{TP} + \text{FP})(\text{TP} + \text{FN})(\text{TN} + \text{FP})(\text{TN} + \text{FN})}}}}
\label{evalmcc}
\end{equation}
where TP represents correctly detected vulnerabilities, TN represents correctly detected non-vulnerabilities, FP represents incorrectly detected vulnerabilities, and FN represents incorrectly detected non-vulnerabilities.

The Coefficient of Variation (CV) is a statistical measure used to determine the dispersion of data points in a dataset relative to its mean. It is particularly valuable when comparing the variability of datasets with different means. The CV is calculated using the equation:

\begin{equation}
\footnotesize
\hspace{2.5cm}
\text{CV} = \frac{\sigma}{\mu}
\label{evalmcc}
\end{equation}
where \(\sigma\) represents the standard deviation and \(\mu\) denotes the mean of the dataset. 

A higher CV indicates a greater level of dispersion within the data distribution, reflecting more variability relative to the mean. $\mathcal{P}_\text{vul}$, $\mathcal{R}_\text{vul}$, $F_1$, and FPR range from 0 to 1, with higher values indicating better performance of a classifier. MCC ranges from -1 to +1, with higher values indicating better performance of a classifier. We use \textbf{percentage values} ($\%$) to highlight the differences between results.

\section{Results and Analysis}
\label{sec:analyis}
This section analyzes the experimental results to address the research questions.

\begin{table*}[htbp]
\caption{Results of DL model comparison}
\scriptsize
\centering
\begin{threeparttable}
\label{tab: compare diff model}
\renewcommand\arraystretch{1.5}

\begin{tabular}{c|lllll|lllll|lllll}
\hline

\rule{0pt}{6.5pt}
\multirow{2}*{\textbf{Project}} & \multicolumn{5}{c}{\textbf{Linevul}} & \multicolumn{5}{c}{\textbf{Devign}} & \multicolumn{5}{c}{\textbf{Sysevr}}\\
\cline{2-16}
& \textbf{$\mathcal{P}_\text{vul}$} & \textbf{$\mathcal{R}_\text{vul}$} & \textbf{$F_1$} & FPR & MCC
& \textbf{$\mathcal{P}_\text{vul}$} & \textbf{$\mathcal{R}_\text{vul}$} & \textbf{$F_1$} & FPR & MCC
& \textbf{$\mathcal{P}_\text{vul}$} & \textbf{$\mathcal{R}_\text{vul}$} & \textbf{$F_1$} & FPR & MCC \\
\hline
\hline
\textbf{Chrome}
&\textbf{40.4} & 73.3 & \textbf{52.1} & \textbf{28.4} & \textbf{37.6}
&29.3 & \textbf{85.5} & 43.7 & 54.0 & 26.1
&27.7& 56.8& 37.2 & 39.0 & 14.6\\
\textbf{Android}
&\textbf{34.6} & \textbf{86.2} & \textbf{49.3} & \textbf{41.4} & \textbf{36.1}
&31.7 & 85.5 & 46.2& 46.7 & 31.3
&29.4& 80.3 & 43.1 & 48.7 & 25.5\\
\textbf{Linux}
&\textbf{57.1}  & \textbf{76.4} & \textbf{65.4} & \textbf{13.9} & \textbf{56.4}
&48.8 & 66.3 & 56.3 & 16.9 & 44.4
&27.7 & 22.6 & 24.9 & 14.4 & 08.8\\
\textbf{Qemu}
&\textbf{84.2} & 55.1 & \textbf{66.7} & \textbf{01.9} & \textbf{63.9}
&52.8 & \textbf{65.5} & 58.5 & 10.7 & 50.3
&28.6 & 10.0 & 14.8 & 04.4 & 09.0\\
\hline
\end{tabular}
\end{threeparttable}
\end{table*}
\subsection{RQ1: Selection of DL Models}
\label{rq1}

We conducted experiments on four large-scale projects to investigate which category of DL model is suitable for \DLAP. The results are provided in \Cref{tab: compare diff model}, which reveal that using Linevul outperforms using others in most datasets and metrics. For instance, in the Chrome dataset, \DLAP with Linevul achieves the highest MCC of 37.6\%, surpassing Devign's 26.1\% and Sysevr's 14.6\%. This finding is consistent in the Linux dataset, where it secures an MCC of 56.4\%, compared to 44.4\% and 8.8\% for Devign and Sysevr, respectively. Furthermore, the precision and F1 scores of Linevul are notably higher across the datasets, underscoring its robustness in identifying vulnerabilities with greater accuracy and fewer false positives, as evidenced by its lower FPR in datasets. Overall, using Linevul surpasses using Devign by an average of 7.2\% and 10.5\% on the comprehensive evaluation metrics F1 and MCC, respectively. It also outperforms integrating Sysevr by an average of 28.4\% and 34.0\% on the same metrics. This demonstrates that Linevul has superior adaptability and generalizability when integrated into LLMs compared to other DL models. These results indicate the effectiveness of integrating Linevul into \DLAP to detect vulnerabilities, especially its superior $F_1$, which implies a higher likelihood of detecting actual vulnerabilities. Its MCC, a critical indicator of the quality of binary classifications, shows the ability of \DLAP with Linevul to solve extremely imbalanced datasets.

\begin{figure*}[htbp]
\scriptsize
\centering
 \includegraphics[width=0.8\textwidth]{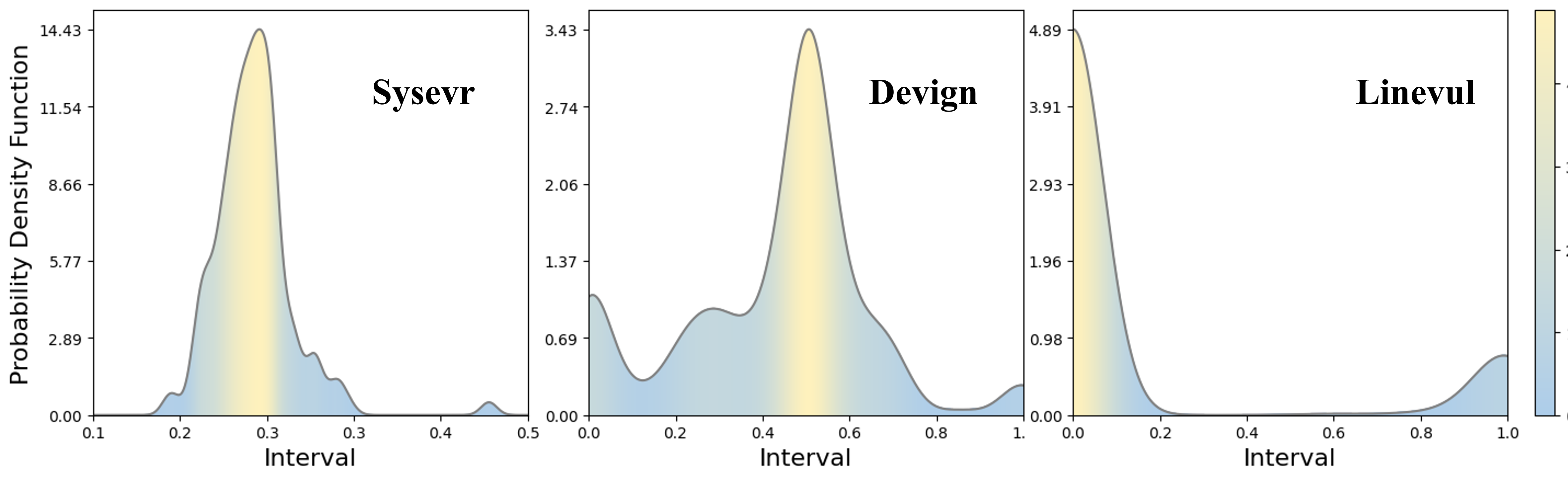}
    \vspace{-2.0ex}
\caption{Distribution of probability density (Sysevr, Devign, and Linevul) on Google project}
    \vspace{-4.0ex}
\label{fig: distrib}
\end{figure*}

\begin{table}[htbp]
\scriptsize
\caption{$CV$ of DL model comparison}
\label{tab: compare CV}
\renewcommand\arraystretch{1.5}
\begin{tabular}{c|llll|l}
\hline
DL model  & Chrome & Android & Linux & Qemu & average\\
\hline
\hline
Sysevr& 0.1 &1.2 & 0.4 & 0.02& 0.43\\
Devign& 0.5 & 1.2 & 2.0 & 2.0& 1.4\\
Linevul& \cellcolor{gray2}\textbf{2.4}& \cellcolor{gray2}\textbf{2.6} &\cellcolor{gray2}\textbf{2.5}& \cellcolor{gray2}\textbf{3.3}& \cellcolor{gray2}\textbf{2.7}\\
\hline
\end{tabular}
\end{table}
To further distinguish which DL model is more suitable as a plug-in for DLAP, we also analyze the intermediate output (detection probability) of the DL model. \Cref{tab: compare CV} presents the variability in the performance of different DL models across several software projects. The Linevul model, highlighted in gray and bold, displays the highest $CV$. By comparing and analyzing probability density distribution plots \Cref{fig: distrib} and $CV$ \Cref{tab: compare CV} in the largest project dataset Google, we noticed that Linevul displays a more discrete data distribution when compared to the other models. This unique discrete detection distribution property facilitates LLM generation with implicit fine-tuning for downstream detection tasks more effectively. 


\begin{tcolorbox}[width=\linewidth, boxrule=1.0pt, left=2pt, right=2pt, top=2pt, bottom=2pt, colback=gray!0]
 \textbf{Summary for RQ1:}
Linevul achieves the best performance among the three types of DL models. Moreover, Linevul results in the most discrete detection probability, indicating its prediction has the highest confidence so that it can stimulate LLMs best. Therefore, we select Linevul as the driver of \DLAP to conduct the follow-up experiments. 

\end{tcolorbox}

\subsection{RQ2: Comparision with Other Prompting Frameworks}\label{rq2}
\begin{table*}[htbp]

\scriptsize
\centering
\begin{threeparttable}
\caption{Results of prompting framework comparison}
\label{tab: baseline comparison}
\renewcommand\arraystretch{1.5}
\begin{tabular}{c|p{0.32cm} p{0.32cm} p{0.36cm} p{0.36cm} p{0.46cm}|p{0.32cm} p{0.32cm} p{0.36cm} p{0.36cm} p{0.46cm}|p{0.32cm} p{0.32cm} p{0.36cm} p{0.36cm} p{0.46cm}|p{0.32cm} p{0.32cm} p{0.36cm} p{0.36cm} p{0.46cm}}
\hline
\rule{0pt}{6.5pt}
\multirow{2}*{\textbf{Framework}} & \multicolumn{5}{c}{\textbf{Chrome}} & \multicolumn{5}{c}{\textbf{Android}} & \multicolumn{5}{c}{\textbf{Linux}}  & \multicolumn{5}{c}{\textbf{Qemu}}\\
\cline{2-21}
& \textbf{$\mathcal{P}_\text{vul}$} & \textbf{$\mathcal{R}_\text{vul}$} & \textbf{$F_1$} & FPR & MCC
& \textbf{$\mathcal{P}_\text{vul}$} & \textbf{$\mathcal{R}_\text{vul}$} & \textbf{$F_1$} & FPR & MCC
& \textbf{$\mathcal{P}_\text{vul}$} & \textbf{$\mathcal{R}_\text{vul}$} & \textbf{$F_1$} & FPR & MCC
& \textbf{$\mathcal{P}_\text{vul}$} & \textbf{$\mathcal{R}_\text{vul}$} & \textbf{$F_1$} & FPR & MCC \\
\hline
\hline
\textbf{\textbf{P}$_{\text{Rol}}$}
&{24.4} & 07.2 & 11.1 & \cellcolor{gray2}\textbf{05.8} & 02.3
&22.5 & 06.4 & 10.0 & 05.6 & 01.3
&22.4 & 06.6 & 10.2 & 05.6 & 01.7
&22.2 & 06.9 & 10.5 & 04.4 & 04.2\\
\textbf{\textbf{P}$_{\text{Aux}}$}
&22.7 & 54.6 & 32.1 & 48.6 & 04.8
&21.8 & 63.4 & 32.5 & 58.3 & 04.2
&24.6 & 70.2 & 36.5 & 52.6 & 14.1
&19.3 & 55.2 & 28.6 & 42.1 & 09.5\\
\textbf{\textbf{P}$_{\text{Cot}}$}
&16.8 & 05.4 & 08.1 & 07.0 & 02.6
&31.6 & 03.1 & 05.7 & \cellcolor{gray2}\textbf{01.7} & 04.0
&30.7 & 08.0 & 12.7 & \cellcolor{gray2}\textbf{04.4} & 06.5
&64.7 & 38.0 & 47.8 & 03.8 & 43.0\\
\textbf{GRACE}
&32.6 & 37.5 & 32.6 & 80.2 & 11.2
&25.0 & 82.6 & 38.4 & 74.0 & 08.5
&25.0 & 76.0 & 37.6 & 76.0 & 02.0
&17.1 & \cellcolor{gray2}\textbf{93.1} & 28.9 & 82.4 & 10.6\\
\textbf{DLAP}
&\cellcolor{gray2}\textbf{40.4} & \cellcolor{gray2}\textbf{73.3} & \cellcolor{gray2}\textbf{52.1} & 28.4 & \cellcolor{gray2}\textbf{37.6}
&\cellcolor{gray2}\textbf{34.6} & \cellcolor{gray2}\textbf{86.2} & \cellcolor{gray2}\textbf{49.3} & 41.4 & \cellcolor{gray2}\textbf{36.1}
&\cellcolor{gray2}\textbf{57.1}  & \cellcolor{gray2}\textbf{76.4} & \cellcolor{gray2}\textbf{65.4} & 13.9 & \cellcolor{gray2}\textbf{56.4}
&\cellcolor{gray2}\textbf{84.2} & 55.1 & \cellcolor{gray2}\textbf{66.7} & \cellcolor{gray2}\textbf{01.9} & \cellcolor{gray2}\textbf{63.9}\\
\hline
\end{tabular}
\end{threeparttable}
\end{table*}

Due to cost constraints associated with OpenAI API calls, we employed the GPT-3.5-turbo-0125 model for vulnerability detection. Table \ref{tab: baseline comparison} illustrates the performance comparison between the GPT model using the baseline prompting framework and the \DLAP approach. The performance of each framework is evaluated based on five metrics: Precision ($\mathcal{P}_\text{vul}$), Recall ($\mathcal{R}_\text{vul}$), F1 Score ($F_1$), False Positive Rate (FPR), and Matthews Correlation Coefficient (MCC). DLAP consistently outperforms the other frameworks across nearly all metrics and datasets. Specifically, DLAP achieves the highest Precision, Recall, F1 Score, and MCC values, showcasing its superior ability to accurately identify vulnerabilities with minimal false positives. For instance, in the Chrome dataset, DLAP's Precision of 40.4\% and Recall of 73.3\% significantly surpass those of the next best framework, GRACE. Furthermore, DLAP's exceptional performance is highlighted by its F1 Score, reaching up to 52.1\% in Chrome, 49.3\% in Android, 65.4\% in Linux, and an impressive 66.7\% in Qemu, which are higher compared to the baseline frameworks. In terms of FPR, DLAP demonstrates a moderate FPR across the datasets. Despite that FPR is not the lowest on the Chrome, Android, and Linux datasets when compared to the baselines of \textbf{P}$_{\text{Rol}}$ and \textbf{P}$_{\text{Cot}}$, \DLAP is far superior to them on the $F_1$ and MCC. Therefore, \DLAP's overall effect exceeds the baseline frameworks.

In particular, \DLAP's MCC values, which indicate the quality of binary classifications, significantly exceed those of the other methods, such as 37.6\% in Chrome and 63.9\% in Qemu, further establishing its superior performance in the task of vulnerability detection using LLMs. Our framework consistently surpasses the top baseline in terms of the MCC indicator, which, based on the nature of the MCC correlation coefficient, suggests that our predictions more accurately reflect the actual distribution and indicates \DLAP is superior to the baselines in the generalization performance of large data sets.

Overall, the analysis reveals that DLAP not only excels in identifying vulnerabilities with high precision and recall but also maintains a low false positive rate and achieves outstanding overall performance as evidenced by its F1 Scores and MCC values. This demonstrates DLAP's exceptional effectiveness in harnessing the power of LLM for the critical task of vulnerability detection, which outperforms the capabilities of other prompting frameworks.

\begin{tcolorbox}[width=\linewidth, boxrule=1.0pt,
  left=2pt, right=2pt, top=2pt, bottom=2pt, colback=gray!0]
 \textbf{Summary for RQ2:} 
\DLAP's overall performance is superior to other prompting frameworks, which is evidenced by its exceptional MCC scores and higher values in  \textbf{$\mathcal{P}_\text{vul}$}, \textbf{$\mathcal{R}_\text{vul}$}, and \textbf{$F_1$}.
\end{tcolorbox}

\subsection{RQ3: Prompting vs. Fine-tuning}\label{rq3}
 
\Cref{tab: compare to FT} shows that fine-tuning an LLM on a large project has a higher $F_1$ than \DLAP. However, on a small project with imbalanced data, \DLAP performs better. In particular, LLMs can not fine-tuned on Qemu because the project has a small amount of data. In contrast, \DLAP gets the distribution characteristics of small samples and hence can achieve better performance. 
In addition, fine-tuning an LLM requires stopping the model and retraining it before using it, whereas \DLAP does not need to be removed for retraining during its use. It is used as a plug-in to access an LLM in real time to augment the vulnerability detection capability of LLMs. Besides, the comparison of computational cost between \DLAP and LoRA fine-tuning is shown on \Cref{tab: compare to cost}. It is clear that fine-tuning a 13B LLM requires close to 40GB of graphics memory and a lot of time. In contrast, \DLAP can select a small DL model and train it to fit the target data in less than one hour.

\begin{table}[htbp]
\scriptsize
\centering
\caption{Results of performance comparison with fine-tuning}
\label{tab: compare to FT}
\renewcommand\arraystretch{1.3}
\begin{threeparttable}
\begin{tabular}{p{0.8cm}|p{0.23cm} p{0.23cm} p{0.26cm} p{0.26cm} p{0.42cm}|p{0.23cm} p{0.23cm} p{0.26cm} p{0.26cm} p{0.42cm}}
\hline
\multirow{2}*{\textbf{Dataset}} 
& \multicolumn{5}{c}{\textbf{Fine-Tuning Vicuna-13B}}
& \multicolumn{5}{c}{\textbf{\DLAP}}  \\
\cline{2-11}
& \textbf{$\mathcal{P}_\text{vul}$} & \textbf{$\mathcal{R}_\text{vul}$} & \textbf{$F_1$} & FPR & MCC
& \textbf{$\mathcal{P}_\text{vul}$} & \textbf{$\mathcal{R}_\text{vul}$} & \textbf{$F_1$} & FPR & MCC \\
\hline
\hline
\textbf{Chrome}
&91.4 & 74.4 & 82.0 & 01.8 & 78.6
&40.4 & 73.3 & 52.1 & 28.4 & 37.6\\
\textbf{Android}
&67.0 & 35.8 & 46.7 & 04.5 & 40.4
&34.6 & 86.2 & 49.3 & 41.4 & 36.0\\
\textbf{Linux}
&96.4 & 55.4 & 70.3 & 00.5 & 68.9
&57.1 & 76.4 & 65.4 & 14.0 & 56.4\\
\textbf{Qemu}
&99.9 & 06.7 & 12.1 & 00.1 & 23.4
&84.2 & 55.2 & 66.7 & 01.9 & 63.9\\
\hline
\hline
\textbf{Total}
&\textbf{88.7} & 43.0 & 52.8 & \textbf{01.2} & \textbf{52.8}
&54.1 & \textbf{72.8} & \textbf{58.4} & 21.4 & 48.5\\
\hline
\end{tabular}
\end{threeparttable}
\end{table}

\begin{table}[htbp]
\scriptsize
\centering
\caption{Reuslts of computational cost comparison}
\renewcommand\arraystretch{1.3}
\label{tab: compare to cost}
\begin{threeparttable}
\begin{tabular}{p{0.8cm}|p{0.5cm} p{0.5cm} l|p{0.5cm} p{0.5cm} l}
\hline
\multirow{2}*{\textbf{Dataset}} 
& \multicolumn{3}{c}{\textbf{Fine-Tuning}}
& \multicolumn{3}{c}{\textbf{\DLAP}}\\
\cline{2-7}
& \textbf{M}(MB) & \textbf{T}(h) & \textbf{GPU}(GB) 
& \textbf{M}(MB) & \textbf{T}(h) & \textbf{GPU}(GB) \\
\hline
\hline
\textbf{Chrome}
& 5.1 & 11.1 & 31.2
& 3.6 & 0.8 & 6.3\\
\textbf{Android}
& 4.9 & 4.2 & 30.3
& 4.3 & 0.5 & 5.5\\
\textbf{Linux}
& 4.9 & 5.5 & 30.3
& 3.8 & 0.4 & 5.5\\
\textbf{Qemu}
& 4.8 & 1.3 & 28.7
& 0.9 & 0.3 & 2.8\\
\hline
\end{tabular}
\end{threeparttable}
\end{table}

\Cref{eqIFT} (Cf. \Cref{sec: framework-ICL}) indicates that DL model training information changes the relaxed attention of the LLM. This results in an implicit fine-tuning for the LLM to adapt to a specific detection task~\cite{dai2023can}. Whether through fine-tuning or In-Context Learning (ICL), the extent to which a model is fine-tuned reflects its ability to adapt to the target task, serving as a crucial factor in stimulating LLMs to perform well.
According to the detection result shown in \Cref{tab: compare to FT}, \DLAP performs well in approximating fine-tuning on performance evaluation metrics. 

\begin{figure}[htbp]
\label{fig: compare distribution}
\centerline{\includegraphics[width=0.5\textwidth]{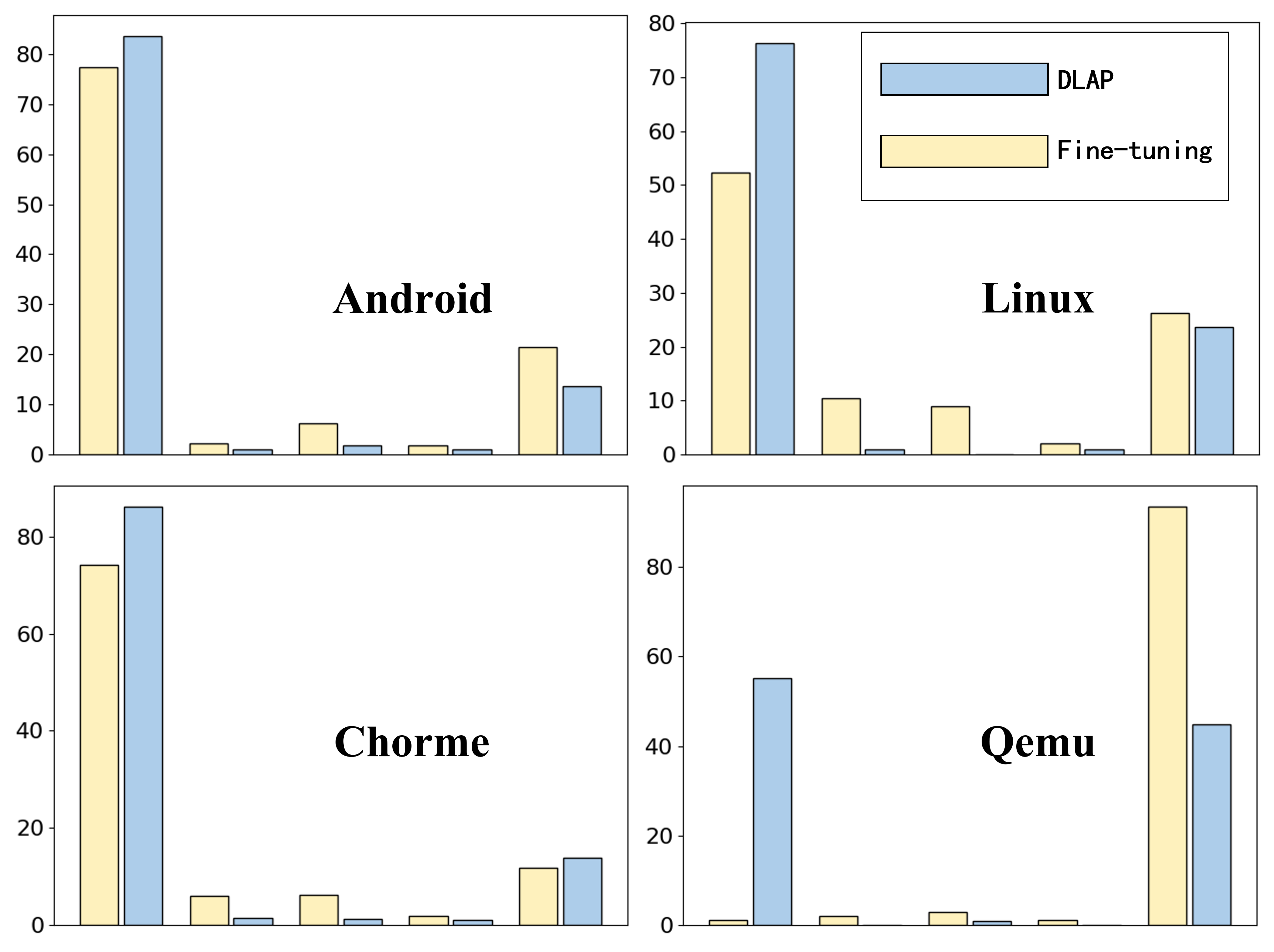}}
    \vspace{-1.0ex}
\caption{Predictive distributions resulted by \DLAP and fine-tuning}
    \vspace{-1.0ex}
\end{figure}

To further explain what mechanism induces LLM to produce implicit fine-tuning and achieve good performance on the target task, we extract the attention layer from the fine-tuned local LLM to calculate the probability for each detection category.
Subsequently, we gather the ICL outputs by the LLM with \DLAP to calculate the probability for each detection category.
The probability distribution of the different classes indicates the degree of fine-tuning of the model.
\Cref{fig: compare distribution} shows that probability distributions between fine-tuning and \DLAP are similar. The same distribution explains that \DLAP enables implicit fine-tuning at a reduced cost. 

\begin{figure*}[htbp]
    \begin{center}
     \hspace{-2cm}

    \includegraphics[width=1.0\textwidth]{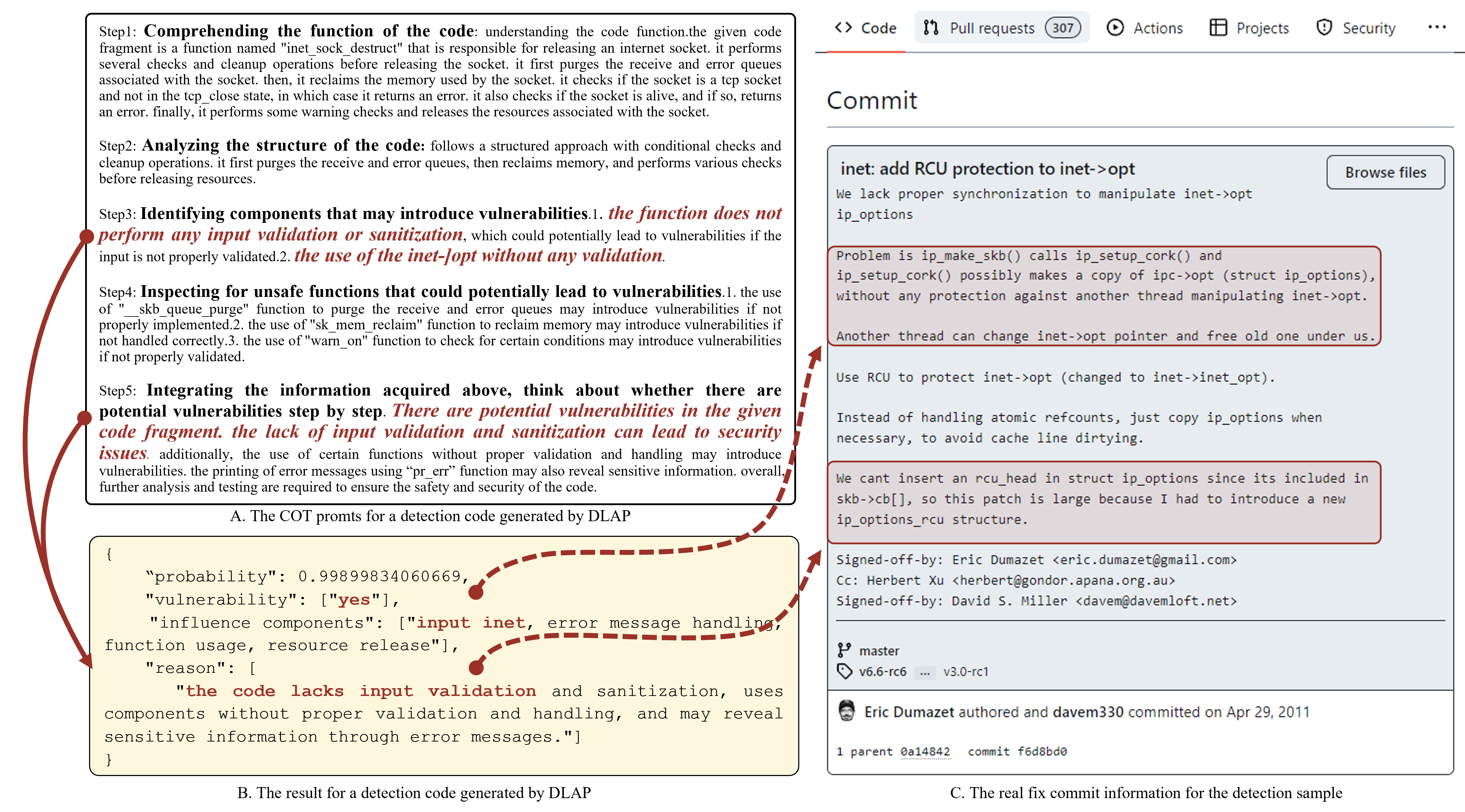}
    \vspace{-4.0ex}
    \end{center}
    \captionsetup{justification=raggedright,singlelinecheck=false} 
    \caption{An example of applying \DLAP to repair Linux code issues (commit id: f6d8bd0)}
    \vspace{-4.0ex}
\label{fig: realissue}
\end{figure*}

In comparison with fine-tuning, \Cref{fig: realissue} shows a real example of using \DLAP to detect vulnerabilities in Linux. The outcomes, which are easily understandable to developers, closely match the records from the actual issue fix commit. In contrast, the output from a fine-tuned LLM is limited to simple `yes' or `no' responses.  \DLAP's results are more comprehensible to developers than those from fine-tuning alone.

\begin{tcolorbox}[width=\linewidth, boxrule=1.0pt, left=2pt, right=2pt, top=2pt, bottom=2pt, colback=gray!0]
\textbf{Summary for RQ3:} Although the overall performance of \DLAP is slightly lower than that of fine-tuning, their predictive distributions are similar. This result shows that \DLAP prompts LLMs to produce effective implicit fine-tuning with performance comparable to that of explicit fine-tuning but at a significantly lower cost.
\end{tcolorbox}

\section{Discussion}
\label{sec:discussion}
In this section, we discuss the DL model selection for \DLAP and \DLAP's potential generalization capability.

\subsection{DL Model Selection for DLAP}

Based on the insights gained from RQ1, DL models with discrete predictive probability density distributions for the data are more suitable as an integrated plug-in for \DLAP. Additionally, we have observed the effectiveness of a DL model as an LLM prompt model. We have discovered that its utility significantly improves when it exhibits discrete data with the highest value of $CV$. Moreover, our experiments have highlighted the exceptional performance of Transformer-based models in driving the LLM. This advantage could be attributed to the architectural resemblance between Transformer models and the design architecture of the LLM. The similarity in their structures allows for seamless integration, enabling the attention-layer parameters derived from Transformer models to play a pivotal role in facilitating implicit fine-tuning within the LLM.

By leveraging these attention-layer parameters, the LLM dynamically adjusts and refines its internal mechanisms, implicitly adapting itself to the nuances and intricacies of different downstream tasks. This implicit fine-tuning process empowers the LLM to generate more accurate and contextually relevant responses, thereby enhancing its overall performance in various application scenarios.

In summary, our experiments have revealed the crucial roles played by both the varied conformity of the DL model and the resemblance between Transformer models and the architecture of LLMs. These factors, combined with the implicit fine-tuning facilitated by attention-layer parameters, enable the LLM to excel in adapting to and fulfilling the requirements of diverse downstream tasks.

\subsection{Generalization Capability of \DLAP}
The \DLAP framework effectively stimulates LLMs to implicitly fine-tune themselves for other software development tasks. By integrating existing static tools and deep learning models, \DLAP is applied to a variety of ASAT tasks. This adaptation simplifies the process of adopting \DLAP to deal with new challenges. We introduce two scenarios that may extend the applicability of \DLAP.

Automated identification of affected libraries from vulnerability data is an ASAT task that requires figuring out which libraries in software are related to each of the reported vulnerabilities in open vulnerability report sets (\eg NVD, CVE). The task is formulated as an extreme multi-label learning~\cite{haryono2022automated, chen2020automated}. First, \DLAP constructs a sufficient vulnerability description database and combines them with libraries known to be affected by the reported vulnerabilities as a COT template library for affected libraries identification. The known affected libraries are collected to train a DL model. Then, existing static tools (fastXML\footnote{\url{https://github.com/fastXML/fastXML}}) and the DL model are used to generate paramilitary results for the XML library list in the project. Finally, by combining the results as a key to query the COT template library, COT prompts can be built to augment the LLMs, and the identification of libraries from vulnerability data may be more accurate.

Code smell detection is an ASAT task that prevents software from technical debts. Code smell detection based on DL models is a multi-classification detection task comprised of several binary classification models, each designed to detect a specific category of code smell~\cite{lewowski2022far,pecorelli2019role}. Utilizing \DLAP requires the creation of a comprehensive reference library of code smells and a high-quality coding standards library. Then \DLAP uses the static tools (checkstyle\footnote{\url{https://checkstyle.org}}) and DL models for specific detection projects. Like the progress mentioned in this paper, employing the DL model augments the LLMs with prompts for the specific project code smell detection.
\DLAP is utilized in other ASAT tasks which need to combine DL models with LLMs to improve the performance of LLMs in the target tasks.

\section{Threats to Validity}
\label{sec:ttv}
This section analyzes possible threats to validity~\cite{zhou2016map} and our efforts to mitigate their impacts.

\textbf{Internal Validity.}
The efficacy of the \DLAP relies on its core component, the DL models. While it is permissible for these DL models to result in judgment biases for input code, a completely erroneous DL model that is either irrelevant or detrimental to the task can severely undermine the performance of the DLAP. Therefore, when employing the DLAP, it is crucial to select DL models that are capable of addressing the specific objectives of the task to effectively augment the performance of LLMs. Besides, due to the close source nature of certain LLMs (GPT-3.5-turbo), their internal structures and the specific fine-tuning methods they employ remain unknown. Therefore, for our experiments, we use an open-source LLM (Llama-13b), for comparative fine-tuning studies.

\textbf{Construct Validity.}
The relaxed attention of the LLMs changes under the stimulation of \DLAP according to~\Cref{eqIFT}. We define this stimulation as the implicit fine-tuning of LLMs caused by \DLAP to adapt to the feature of the target project. Because of the limitations of observing the internal output of LLMs, we can not strictly demonstrate that the stimulation produces gradient descent optimization loss on the target classification task. Instead of performing a demonstration mathematically, we show some of our results with advantages and some of the intermediate outputs of some contrasts showing fine-tuning, validating the existence of the implicit fine-tuning mechanism in the way of experimental data. These visualized experimental results eliminate the construct validity of this paper to some extent. 

\textbf{External Validity.} 
For the verification of \DLAP, the final effect of our template needs to drive LLM to complete the vulnerability detection, and the performance of this task is used as an evaluation to measure the effectiveness of our method. So when the LLM is different from the LLM selected in this experiment, the results of using \DLAP will be different. We identify the choice of LLM as an external validity threat to this work. Considering both cost and model performance, we chose the least expensive model of the current state-of-the-art LLMs, GPT-3.5-turbo-0125. By using the best model, we make the best use of \DLAP. we provide specific model selection, which ensures that other work reproduces the same level of improvement when using \DLAP.

\section{Conclusion}
\label{sec:conclusion}
In this paper, we propose \DLAP, a bespoke prompting framework for ASAT tasks that has superior and stable performance in software vulnerability detection tasks with results easily understandable to developers. Experiments show the effectiveness of augmenting LLMs by DL models to stimulate adaptive implicit fine-tuning. This progress prompts LLMs to exceed both state-of-the-art DL solutions and LLMs with alternative prompting frameworks in vulnerability detection. Through experiments, we also find that the pre-trained knowledge of LLMs combines the outputs of all parts of \DLAP to achieve good performance. In the future, we will utilize \DLAP in more ASAT tasks to explore how \DLAP is generalized to the other tasks.

\bibliographystyle{cas-model2-names}
\bibliography{references}

\section*{Appendix}
\label{proveICL}
The appendix analyzes the composition and causes of implicit fine-tuning.
The implicit fine-tuning process is described as follows.  
we sets $ \mathcal{C}(x)=COT(x)$ as the input representation for the context. It sets the target task data input as $x$, and $ \mathcal{P}(x)=ICL(x) $ as the input representation for \DLAP prompts query. $W_{Q}, W_{K}, W_{V} $ are the projection matrices for computing the attention queries.  $\mathbf{q} = W_{Q}$ is the attention query vector. In the fine-tuning process of \DLAP, the attention of LLMs is represented as the \Cref{eq1}
\begin{equation}
\footnotesize
    \begin{aligned}
        \mathcal{A}(\mathbf{q}) & = \operatorname{Attention}(V, K, \mathbf{q}) \\
        = & W_{V} [\mathcal{P}(x); C(x)] \operatorname{softmax} \left( \frac{\left( W_{K} [\mathcal{B}(x); C(x)] \right)^T \mathbf{q}}{\sqrt{d}} \right) 
    \end{aligned}
\label{eq1}
\end{equation}
where $W_{Q}, W_{K}, W_{V} $ are the projection matrices for computing the attention queries. To simplify this process, we eliminate the nonlinear function $\operatorname{softmax}$ and related scaling factor $\sqrt{d}$ to facilitate the analysis of the attention changes in this process itself. We obtain an approximate relaxed linear attention, \Cref{eq2}.
\begin{equation}
\footnotesize
    \begin{aligned}
    \mathcal{A}(\mathbf{q}) & \approx 
    W_{V} [\mathcal{P}(x); C(x)] \left( W_{K} [\mathcal{P}(x); C(x)] \right)^T \mathbf{q} \\
     & = W_{V} C(x) \left( W_{K} C(x) \right)^T \mathbf{q} + W_{V} \mathcal{P}(x) \left( W_{K} \mathcal{P}(x) \right)^T \mathbf{q} \\
    & = \widetilde{\mathcal{A}}(\mathbf{q})
    \end{aligned}
\label{eq2}
\end{equation}

We define the context (information from COT library) prompts as the initial parameters $W_{\text{init}}$ that need to be updated by the attention layer in \Cref{eq3}. 
\begin{equation}
\footnotesize
 \begin{aligned}
W_{\text{init}}=W_v[\mathcal{C}(x)]\cdot W_K [\mathcal{C}(x)]^T\cdot \mathbf{q})
    \end{aligned}
\label{eq3}
\end{equation}
According to research~\cite{dai2023can}, we reverse use the dual form of the attention of transformer derived by them. Therefore, the adaptive implicit fine-tuning of LLMs stimulated by \DLAP for specific projects are written as the \Cref{eq2}
\begin{equation}
\footnotesize
 \begin{aligned}
 \widetilde{\mathcal{A}}(\mathbf{q}) & = W_{\text{init}} \mathbf{q} + W_{V} [\mathcal{P}(x)] \left( W_{K} [\mathcal{P}(x)] \right)^T \mathbf{q} \\
        = & W_{\text{init}} \mathbf{q} + \operatorname{LinearAttn} \left( W_{V} [\mathcal{P}(x)], W_{K} [\mathcal{P}(x)], \mathbf{q} \right) \\
        = & W_{\text{init}} \mathbf{q} + \sum_i \left( (W_{V} \textbf[\mathcal{P}(x)]_i) \otimes \left( W_{K} \textbf[\mathcal{P}(x)]_i \right) \right) \mathbf{q} \\
        = & W_{\text{init}} \mathbf{q} + \Delta W_{\mathcal{P}(x)} \mathbf{q} \\
        = & \left( W_{\text{init}} + \Delta W_{\mathcal{P}(x)} \right) \mathbf{q}
    \end{aligned}
\label{eq4}
\end{equation}
Through \label{eq4}, we conclude that the relaxed attention mechanism is influenced by the prompt $\mathcal{P}$.

\end{document}